\documentclass[sn-mathphys,Numbered]{sn-jnl}% Math and Physical Sciences Reference Style
\usepackage{graphicx, tikz}%
\usepackage{multirow}%
\usepackage{amsmath,amssymb,amsfonts}%
\usepackage{mathrsfs}%
\usepackage[title]{appendix}%
\usepackage{xcolor}%
\usepackage{textcomp}%
\usepackage{manyfoot}%
\usepackage{booktabs}%
\usepackage{algorithm}%
\usepackage{algorithmicx}%
\usepackage{algpseudocode}%
\usepackage{listings}%
\usepackage{mathtools, enumerate, hyperref, enumerate, dsfont, csquotes, paralist, bbm, url, textcomp, doi}
\usepackage[nice]{nicefrac}
\allowdisplaybreaks
 
\newcommand{\RR}{{\mathbb R}}

\newcommand{\cF}{\mathcal{F}}

\newcommand{\cD}{\mathcal{D}}
\newcommand{\eps}{\epsilon}
\newcommand{\E}{{\mathbb E}}

\newcommand{\argmax}{\text{argmax}}

\newcommand{\TT}{{\mathcal T}}

\def\b0{{\bf 0}}
\def\b1{{\bf 1}}

\usepackage{amsthm}
\newtheorem{theorem}{Theorem}
\newtheorem{lemma}{Lemma}
\newtheorem{corollary}{Corollary}
\newtheorem{conj}{Conjecture}
\newtheorem{claim}{Claim}

\begin{document}

\def\UrlBreaks{\do\/\do-}

\title[Towards an Optimal Contention Resolution Scheme for Matchings]{Towards an Optimal Contention Resolution Scheme for Matchings\footnote[3]{A version of this paper was published in the proceedings of the conference IPCO in 2023 \cite{NV23}. This version contains all omitted proofs, discussion of the relationship of this paper with van der Waerden's conjecture, and an application of our contention resolution scheme to a combinatorial allocation problem. The second author is supported by NSF Award 2127781.}}


\author*[1]{\fnm{Pranav} \sur{Nuti}}\email{pranavn@stanford.edu}
\equalcont{These authors contributed equally to this work.}

\author[1]{\fnm{Jan} \sur{Vondrák}}\email{jvondrak@stanford.edu}
\equalcont{These authors contributed equally to this work.}

\affil[1]{\orgdiv{Department of Mathematics}, \orgname{Stanford University}, \orgaddress{\city{Stanford}, \postcode{94305}, \state{CA}, \country{USA}}}

\abstract{In this paper, we study contention resolution schemes for matchings. Given a fractional matching $x$ and a random set $R(x)$ where each edge $e$ appears independently with probability $x_e$, we want to select a matching $M \subseteq R(x)$ such that $\Pr[e \in M \mid e \in R(x)] \geq c$, for $c$ as large as possible. We call such a selection method a $c$-balanced contention resolution scheme.

Our main results are (i) an asymptotically optimal  $\simeq 0.544$-balanced contention resolution scheme for general matchings when $\|x\|_\infty \to 0$, and (ii) a $0.509$-balanced contention resolution scheme for bipartite matchings (without any restriction on $x$). To the best of our knowledge, this result establishes for the first time, in any natural relaxation of a combinatorial optimization problem, a separation between  (i) offline and random order online contention resolution schemes, and (ii) monotone and non-monotone contention resolution schemes.}




\maketitle

\section{Introduction}

Suppose that there are $n$ employees looking for jobs. Each employee likes a random set of jobs which, on average, has cardinality one. $n$ jobs are available in total, and no job is especially popular amongst the employees, though some employees might have a strong preference for some particular jobs. We would like to match the employees to jobs.

We are immediately faced with many natural questions: On average, what fraction of employees can we match to a job they like? Can we match employees to jobs in a fair way, without partially favoring any particular employee? What if no employee has a strong preference for any particular job? Is it easier to match employees if we learn about their preferences all at once, rather than if we learn about them in an online fashion? Our paper provides answers to these questions, through the lens of contention resolution schemes.

Contention resolution schemes aim to solve the following problem:
Given a family of feasible sets $\cF \subset 2^E$ and a random set $R$ sampled from a distribution on $2^E$, how can we choose a feasible subset $I \subseteq R, I \in \cF$, so that each element from $R$ is picked with some guaranteed conditional probability: $\Pr[e \in I \mid e \in R] \geq c$ for some fixed $c>0$ and all $e \in E$? We call such a scheme {\em $c$-balanced}. This condition is a kind of fairness constraint, ensuring every element $e$ has a reasonable chance of making it into $I$.

In this paper, we think about $E$ as the set of edges in a graph, and $\cF$ as the set of matchings of the graph. The constant $c$ is the conditional probability with which we can ensure an edge ends up in the matching $I$ we pick, given it appears in $R$.

A natural assumption on the random set $R$ is that it comes from a product distribution with marginal probabilities $x_e$ such that $x$ is in a polytope corresponding to the family $\cF$ (either the exact convex hull, or a suitable relaxation, depending on the application), i.e., roughly speaking, $R$ on average, is in $\cF$. For matchings on graphs, this corresponds to an assumption that each edge $e$ appears in $R$ independently with probability $x_e$, and the vector $(x_e)_{e \in E}$ belongs to the matching polytope, i.e, is a \textit{fractional matching}.

The formal notion of contention resolution was first investigated as a tool for randomized rounding. In this setting, we have an optimization problem subject to a constraint, and $x$ represents a fractional solution to a relaxation of the problem. Contention resolution is one of the phases of a randomized rounding approach to converting this fractional solution into an integral solution: First we generate a random set $R$, by sampling each element $e$ independently with probability $x_e$, and then we select a subset $I \subseteq R$ which satisfies the desired constraint. The flexibility of the approach enables its wide applicability in combinatorial optimization.

This approach was introduced by Feige \cite{Feige09},  who developed a contention resolution scheme (CRS) for the constraint $\{I: |I| \leq 1\}$ (equivalent to matchings on a star graph), in the context of an application to combinatorial auctions. CRSs were then investigated more systematically in \cite{CVZ14} in the context of submodular optimization. In particular, an optimal $(1-1/e)$-balanced CRS was identified in \cite{CVZ14} for the case where $\cF$ forms a matroid. The $1-1/e$ factor is optimal even for $\cF = \{ I: |I| \leq 1 \}$. 

For applications in submodular optimization involving randomized rounding \cite{CVZ14}, it turns out that an additional property of {\em monotonicity} is often useful: A CRS is called monotone, if for every element $e$, the probability that $e$ is selected from a set $R$ is non-increasing as a function on the sets $R$ containing $e$. However, for many other applications, it is not necessary that a CRS is monotone; in particular it was not needed in Feige's original application in \cite{Feige09}, it is unnecessary for a related application that we present (involving rounding of the {\em Configuration LP} for an assignment problem) in Section~\ref{sec:application}, and it also unnecessary for direct applications such as the one mentioned at the beginning of this paper. The schemes that we develop in this paper are generally {\em non-monotone}. 

Contention resolution has also been studied in online settings (where it has seen applications to prophet inequalities and sequential pricing problems, for example) with either adversarial or random ordering of elements \cite{FSZ16,AW18, LS18,FTWWZ21,PRSW22}. For example, for matroids there is a $1/2$-balanced adversarial order online CRS \cite{LS18}. We do not investigate online contention resolution here, but we should mention that in prior results, random-order online contention resolution schemes (RCRS) are able to match the best known offline results: For matroids, there is a $(1-1/e)$-balanced RCRS, due to an elegant LP duality connection with prophet inequalities \cite{LS18}. 

The situation is more complicated when $\cF$ encodes constraints such as matchings or intersections of matroids, and the optimal factors are generally unknown. The cases of bipartite and general matchings have attracted attention due to their fundamental nature and their frequent appearance in applications. We can think of bipartite matching as an intersection of two matroid constraints. For an intersection of $k$ matroid constraints, there is a $\frac{1}{k+1}$-balanced RCRS \cite{AW18}; in particular, this gives a $\frac13$-balanced RCRS (and hence also an offline CRS with the same factor) for bipartite matchings.

Recent work in both offline and online settings has significantly improved the factor of $\frac13$. In the offline setting, \cite{GL17} gives a $\frac12 (1-e^{-2}) \simeq 0.432$-balanced scheme for general matchings, which can be improved slightly further  \cite{BZ22}. Very interestingly, \cite{BZ22} identifies the optimal monotone scheme for bipartite matchings, which achieves a balancedness of $\simeq 0.476$.  Nevertheless, the optimal CRSs for bipartite and general matchings are still unknown. The primary reason that it seems to be harder to obtain the optimal non-monotone scheme is that decisions on whether an edge should be included in the matching need not be local (i.e., a function of the edge's immediate neighborhood), and it is harder to analyze the behavior of an algorithm that makes non-local decisions. In terms of impossibility results, an upper bound of $\simeq 0.544$ follows from a classical paper of Karp and Sipser \cite{KS81}, as discussed in \cite{GL17}.

In the online setting, the best known CRSs are due to recent results in \cite{MMG2022}: In the random order case, they provide a $0.474$-balanced scheme for general matchings and a $0.476$-balanced scheme for bipartite matchings, and in the bipartite case, they also establish an upper bound of $0.5$. Notably, the $0.474$-balanced scheme is in fact the best known CRS for general matchings, whether offline or online. In the adversarial order case, \cite{MMG2022} gives a $0.344$-balanced scheme for general matchings and a $0.349$-balanced scheme for bipartite matchings.

\subsection{Our results}

To explain our results, we start by formally setting up some notation. Given a graph $G = (V, E)$, a fractional matching is a point $x \in [0,1]^E$ in the matching polytope, i.e., a point in the convex hull of vectors $\b1_M$ for all matchings $M$ in $G$. For a fractional matching $x$, let $x_{uv}$ be the component of $x$ corresponding to the edge $(u, v)$.

The problem we are interested in studying is:
\paragraph{Contention resolution for matchings.} We are given a fractional matching $x$, and a random set $R(x)$ where edges appear independently with probabilities $x_{uv}$. Our goal is to choose a matching $M \subseteq R(x)$ such that for every edge $(u,v)$, 
$$\Pr[(u,v) \in M \mid (u,v) \in R(x)] \geq c.$$
Such a scheme is called $c$-balanced, and we want to find a scheme with $c$ as large as possible. The main questions we ask are:
\begin{enumerate}
\item[(i)] Is there a contention resolution scheme for matchings achieving the upper bound $\simeq 0.544$ of Karp and Sipser?
\item[(ii)] Is there a separation between the optimal $c$ for online and offline contention resolution schemes?
\item[(iii)]  Is there a separation between the optimal $c$ for monotone and non-monotone contention resolution schemes?
\end{enumerate}

In this paper, we prove the following results. The first result, which applies to both bipartite and non-bipartite matchings, is an attempt to answer (i).

\begin{theorem}
\label{thm:0.544}
Suppose that $x$ is a fractional matching such that $\|x\|_\infty \leq \epsilon$. Then, there is a function $f$ (not dependent on $x$) with $\lim_{\epsilon \rightarrow 0} f(\epsilon) = 0$, and a $(\gamma - f(\epsilon))$-balanced contention resolution scheme, where $\gamma \simeq 0.544$ is the impossibility bound of Karp and Sipser.
\end{theorem}

For fractional matchings without any assumption on their $\ell_\infty$ norm\footnote{
It might appear from the work of Bruggmann and Zenklusen (see lemma 7 in \cite{BZ22}) that the assumption of $\|x\|_\infty \leq \epsilon$ should be easy to drop from Theorem~\ref{thm:0.544}. This would be the case if our theorem applied to graphs with parallel edges, which unfortunately, it does not. }, we present an improved CRS in the bipartite case.

\begin{theorem}
\label{thm:0.509}
There is $0.509$-balanced contention resolution scheme for bipartite matchings.
\end{theorem}

This theorem answers questions (ii) and (iii), since the optimal RCRS for bipartite matchings is at most $0.5$-balanced, and the optimal monotone CRS for bipartite matchings is $\simeq 0.476$-balanced. Our theorem thus establishes separations that, to our knowledge, have not been demonstrated in any other natural relaxations of combinatorial optimization problems before.  (Note that for matroids, the known optimal $(1-1/e)$-balanced schemes are monotone.)

Returning to the context we started this paper with, our results establish that we can match more than half of all the employees to jobs they like without partially favoring any particular employee, and in case no employee has a strong preference for any particular job, we can do better, and match 54$\%$ of employees to jobs. This is a significant improvement over what we can do if we learn the employees preferences in an online fashion.

We should also mention here the important concept of a \textit{correlation gap}. Informally, the correlation gap measures how much we might lose while optimizing a function, in the worst case, by assuming that the distributions that define the function are independent rather than correlated. In the context of bipartite matchings, the correlation gap is defined as the minimum possible ratio between $\E[\max \{ \sum_{e \in M} w_e: M \subseteq R(x), \text{ } M \text{ is a matching} \}]$ and $\sum_{e \in E} w_e x_e$, where $x$ is a fractional bipartite matching and $w$ is any vector of weights. By LP duality (see \cite{CVZ14}), Theorem~\ref{thm:0.509} also provides (the best known) lower bound of $0.509$ on the correlation gap for bipartite matchings\footnote{Note that we do not need a monotone contention resolution scheme to establish such a lower bound.}. 

In light of Theorem~\ref{thm:0.544}, we believe that the correlation gap for bipartite (and perhaps even non-bipartite) matchings is indeed the Karp-Sipser bound of $\gamma \simeq 0.544$, and the optimal CRS is $\gamma$-balanced. This conjecture has intriguing connections with van der Waerden's conjecture for the permanent of doubly stochastic matrices; we discuss this in Section~\ref{sec:vanderW}.

\subsection{Our techniques}

Our Theorem~\ref{thm:0.544} follows from an improved and simplified analysis of Karp and Sipser's algorithm \cite{KS81} for constructing matchings. Given a graph $G$, the algorithm selects a random degree 1 vertex (if one exists), and adds the edge adjacent to it to the matching. Then, it deletes all the edges adjacent to the edge just added to the matching, and recurses on the newly obtained graph. While we utilize many of the ideas from Karp and Sipser's paper, our analysis of the algorithm is an improvement in several ways:
\begin{itemize}
    \item We obtain a contention resolution scheme, while Karp and Sipser only compute the expected size of the maximum matching. This yields the somewhat surprising conclusion that Karp and Sipser's algorithm works just as well for weighted matchings as it does for unweighted matchings.
    \item We avoid Karp and Sipser's (technically complicated) use of the so-called differential equation method. We also avoid the use of generating functions, another method used recently to calculate the expected size of the maximum matching in random graphs  \cite{controllability}.
    \item We obtain results for any random graph $R(x)$ constructed from a fractional matching satisfying $\|x\|_\infty \leq \epsilon$, unlike Karp and Sipser who only consider the Erdos-Renyi random graph $G_{n,c/n}$. 
\end{itemize}
Many previous results require that there be some kind of symmetry in the random graph to obtain bounds on the size of the matching. We stress that we do not need to make any such assumption on $R(x)$. 

We do need to assume that $\|x\|_\infty \leq \epsilon$. This assumption is useful because it ensures that the neighbourhood of any particular edge looks like a certain random tree, and we can then restrict ourselves to analyzing the performance of the Karp-Sipser algorithm on this random tree. A closely related assumption (``local weak convergence") has been considered previously in the literature. This assumption, together with recursive distributional equations, is used to formalize various statistical mechanical heuristics regarding matchings in random graphs. Most related to our work is the work of Bordenave, Lelarge, and Salez \cite{BLS2013}. Once again, the advantage of our method is that we obtain a CRS (as opposed to computing the expected size of the maximum matching) and we avoid the use of technically complicated tools.

Compared to \cite{KS81}, we assume that $\sum_{v: (u,v) \in E} x_{uv} \leq 1$ for every vertex $u$ (corresponding to $c=1$ in \cite{KS81}). One could consider the more general constraint $\sum_v x_{uv} \leq c$, but in the context of a contention resolution scheme applied to linear relaxation of a matching problem, the constraint $\sum_v x_{uv} \leq 1$ is the most natural one.

%These improvements come at a cost--we assume that the average degree of each vertex is less than or equal to 1. The theoretical and practical significance of this case, and the importance of contention resolution schemes, make this trade-off a good choice.

Our Theorem~\ref{thm:0.509} requires several new techniques, although the basic idea can be traced back to Karp and Sipser as well: When deciding which edge incident to a vertex we should add to a matching, it is beneficial to pick an edge which is adjacent to a leaf, since it doesn't block us from adding other edges into the matching. It turns out that in general, it is actually better not to follow this rule absolutely (at least in our analysis) but we still pick degree-$1$ edges with significant priority over other edges. 

We present two different schemes using these ideas; the first one is simpler and achieves a factor $\simeq 0.480$ (already establishing the separation between monotone and non-monotone schemes). An interesting feature of this scheme is that it can be implemented as a parallel algorithm with each vertex independently making decisions about whether to include an edge adjacent to it in the matching by looking only at its immediate neighborhood. The best schemes known previously did not have this useful property. Our more complicated scheme achieves a factor $\simeq 0.509$ (thus demonstrating a separation between offline CRSs and RCRSs). 

Both schemes rely on an extended version of contention resolution for choosing 1 element from a possibly {\em correlated distribution}, which we present in Section~\ref{sec:CRS-1}, and the $0.509$-balanced scheme uses the FKG inequality to handle correlations between edges in the final stage. 

Throughout this paper, even though we state our theorems for fractional matchings $x$, we will actually only need to assume that $x$ satisfies the vertex constraints $\sum_{v} x_{uv} \leq 1$. Furthermore, we can always assume that $x$ satisfies $\sum_{v} x_{uv} = 1$ for every $u$. We can achieve this by adding vertices and edges with probabilities such that the edge probabilities at each vertex add up to $1$; this only makes the task of designing a CRS more difficult.

\subsection{Parallels with van der Waerden's conjecture}
\label{sec:vanderW}

Recall the classical van der Waerden's conjecture (in fact a theorem \cite{EGORYCHEV1981299}):

\paragraph{van der Waerden's conjecture} {\em Given any doubly stochastic $n \times n$ matrix $A$ (an array of coefficients $(a_{ij})_{i,j=1}^{n}$ such that $\sum_{i=1}^{n} a_{ij} = 1$ for every $j$ and $\sum_{j=1}^{n} a_{ij} = 1$ for every $i$), the permanent of $A$ is at least $n! / n^n$ (which is the permanent of the matrix where all entries are equal to $1/n$).}

\

Every doubly stochastic matrix $A$ corresponds to a fractional matching $x$ of $K_{n,,n}$ in a natural way: If $u$ is the $i^\text{th}$ vertex on the left, and $v$ is the $j^\text{th}$ vertex on the right, $x_{uv} = a_{ij}$. Therefore, given any doubly stochastic matrix $A$, we can consider a random bipartite graph $R(x)$.
In this interpretation, the permanent of the matrix is the {\em expected number of perfect matchings} in $R(x)$. Hence, we have the following reformulation:

\paragraph{van der Waerden's conjecture, reformulated} {\em The expected number of perfect matchings in $R(x)$, as a function of a fractional bipartite matching $x$, is minimized when $x_{uv}= \frac{1}{n}$ for all edges $(u, v)$.}

\

Our first conjecture can be therefore formulated as follows.

\begin{conj}
The expected size of the maximum matching in $R(x)$, as a function of a fractional bipartite matching $x$, is minimized when $x_{uv}= \frac{1}{n}$ for all edges $(u, v)$. 

In other words, if $A$ is a doubly stochastic matrix and $R$ is a random bipartite graph where the edge between the $i^\text{th}$ vertex on the left and the $j^\text{th}$ vertex on the right appears in $R$ independently with probability $a_{ij}$, then the expected size of the maximum matching in $R$ is minimized when $a_{ij} = 1/n$ for all $(i,j)$. 
\end{conj}

It is well known that the expected size of the maximum matching in $R(x) \subset K_{n,n}$ where $x_{uv} = 1/n$ is $(\gamma + o(1)) n$ where $\gamma = 2 (1-\lambda) - \lambda^2$ and $\lambda = e^{-\lambda}$ ($\lambda \simeq 0.567$ and $\gamma \simeq 0.544$)\cite{GL17} (and we also provide an explanation of this fact in as a consequence of Theorem~\ref{thm:0.544}). Our conjecture is that this is the worst case among all fractional matchings in $K_{n,n}$.

By LP duality between CRSs and the quantities discussed in this section (see \cite{CVZ14}), we would obtain an optimal $\gamma$-balanced CRS for bipartite matchings if the following weighted version of the conjecture were true. 

\begin{conj}
Let $A$ be a doubly stochastic matrix and $W$ a matrix of weights such that $\sum_{i,j=1}^{n} w_{ij} a_{ij} = 1$. Then for a random bipartite graph $R$ where the edge between the $i^\text{th}$ vertex on the left and the $j^\text{th}$ vertex on the right appears in $R$ independently with probability $a_{ij}$,  the expected maximum-weight matching in $R$ is minimized when $a_{ij} = w_{ij} = 1/n$ for all $(i,j)$. 
\end{conj}

An extension of van der Waerden's conjecture which is also known to be true is the following (see e.g.~\cite{Gyires96}, Theorem 4.3): For any $1 \leq k \leq n$, the expected number of $k$-matchings (matchings of $k$ edges) in the random set $R(x)$ as above is minimized again for $x_{uv} = 1/n$ for all edges $(u, v)$. This however does not imply that the {\em probability that a $k$-matching exists}, a quantity useful to consider to establish the first conjecture, is minimized for the same matrix. The thresholds for the existence of a $k$-matching and the expected number of $k$-matchings are not the same. It can be verified by relatively straightforward computations that for $x_{uv} = 1/n$, the expected number of $k$-matchings is large for $k = 0.6 n$, but the probability that a $0.6n$-matching exists is vanishingly small. This is the discrepancy between the existence and expectation thresholds, addressed by the Kahn-Kalai conjecture \cite{KAHN_KALAI_2007}; however, the recent resolution of it  \cite{PP22} does not shed any light on our problem since the general bounds on the gap are logarithmic.

Nevertheless, we believe that these parallels with prominent problems and results in probabilistic combinatorics make our conjectures appealing.

\section{An optimal contention resolution scheme for vanishing probabilities}
%when $\|x\|_\infty \to 0$}

In this section, we will analyze the Karp-Sipser algorithm to prove the following rephrasing and strengthening of Theorem \ref{thm:0.544}.

\begin{theorem}
\label{thm:Karp-Sipser}
There is a function $f:[0,1] \to [0,1]$ with $\lim_{\epsilon \rightarrow 0} f(\epsilon) = 0$, and an  algorithm $\mathsf{A}$ such that for any graph $G=(V,E)$ and any $x \in [0,1]^E$ satisfying $\sum_{v: (u,v) \in E} x_{uv} = 1$  for every vertex $u$, and $\|x\|_\infty \leq \epsilon$: The algorithm $\mathsf{A}$, given the random graph $R(x)$, selects a matching $M \subseteq R(x)$ with the property that
$$ \Pr[(u, v)\in M\mid(u,v) \in R(x)] \geq 2(1-\lambda)-\lambda^2 - f(\epsilon) $$
where $\lambda$ is the unique real root of the equation $\lambda = e^{-\lambda}$. 

Furthermore, this algorithm is optimal in the sense that
$$ \mathbb E[\text{size of maximum matching in } R(x)] \leq \left(1-\lambda-\frac12 \lambda^2 + f(\epsilon) \right) |V|.$$
\end{theorem}

The main idea behind the proof of Theorem \ref{thm:Karp-Sipser} is the following: The connected component of a given edge $(u, v)$ in $R(x)$ looks like a random tree $\TT$ containing $(u, v)$ in which each vertex $w$ has $X_w$ many children, where $X_w$ is a Poisson random variable with mean $1$. This is because the number of neighbours of a vertex in $R(x)$ is the sum of the independent Bernoulli random variables corresponding to the existence of edges adjacent to it in $R(x)$, and Le Cam's theorem tells us that such a sum of many  independent Bernoulli random variables with small mean has a roughly Poisson distribution. The assumption that $\|x\|_\infty \leq \epsilon$ is essential to this step, and there does not seem to be an easy way to drop this assumption using our current approach. We establish this correspondence in Lemma \ref{lemma:randomtree}.

Once we have a correspondence between $R(x)$ and random trees, we turn to finding a contention resolution scheme for random trees. It turns out that the Karp-Sipser algorithm is extremely convenient for this purpose. In sections \ref{sec:Karp-Sipsertrees} and \ref{sec:Karp-Sipserrandomtrees}, we use a recurrence to study the performance of the Karp-Sipser algorithm on random trees, and establish that it selects an edge $(u, v)$ with probability exactly $2(1- \lambda) - \lambda^2$. Finally, we end by establishing the correspondence between the performance of Karp-Sipser's algorithm on $R(x)$ and random trees.

\subsection{The Karp-Sipser Algorithm}
\label{sec:KSalg}

The Karp-Sipser algorithm is a method to select a matching in a graph. Given a graph $G$, the algorithm deletes all the degree 0 vertices, selects a random degree 1 vertex (if one exists), and adds the edge adjacent to it to the matching. Then, it deletes all the edges adjacent to the edge just added to the matching, and recurses on the newly obtained graph $G'$. Once the graph does not contain any degree $1$ vertices, the algorithm proceeds to a second stage to select additional edges; however, we do \textit{not} need the second stage to obtain our result. In the following, we only consider the first stage.

An attractive feature of the first stage of the Karp-Sipser algorithm is that it ``doesn't make any mistakes". This is because for any vertex $v$ of degree 1 in a graph $G$, $G$ has a maximum matching in which $v$ is matched. 

If an edge is deleted by the algorithm, we will say that {\em it disappears}. We also say that a {\em vertex is matched} if an edge incident to it is added to the matching. Before we discuss the analysis of the algorithm, we take a brief detour.

\subsection{Random Trees}

Consider the following method to generate a random tree. Fix two special vertices, $u$ and $v$, and draw an edge between them. We consider $u,v$ to be at depth $0$. In step $i$, for each vertex at the depth $i-1$, independently sample a Poisson random variable with mean 1, and add this many children to the vertex. Stop at step $j$ if there are no vertices at depth $j-1$. Let us call the random tree generated by this process $\TT$.

Since the two subtrees of $u$ and $v$ are independent copies of a Galton-Watson process with $1$ expected child at each node, it is straightforward to prove that this process terminates with probability 1 (e.g., see Theorem 6.1 in \cite{branchingprocessestext}). So it is almost always true that this process produces a finite tree.

The following lemma explains why we care about the process: Up to small errors, it describes the distribution of the connected component containing a given edge $(u,v)$ in $R(x)$. 

\begin{lemma}
\label{lemma:randomtree}
Let $x$ be a fractional matching with $\sum_w x_{vw} = 1$ for every vertex $v$, and $\|x\|_\infty \leq \epsilon$. Let $R(x)$ be the corresponding random graph. Let us condition on $(u,v) \in R(x)$ and define $N((u,v))$ to be the connected component in $R(x)$ containing $(u,v)$. Let $\TT$ be a random tree produced by the process described above and let $T_0$ be any fixed finite realization of the process. Then 
$$ \Big| \Pr_{R(x)}[N((u, v)) =  T_0 \mid(u,v) \in R(x)] - \Pr_{T \sim \TT}[T=T_0] \Big| \leq 3 \epsilon |T_0|^2.$$
\end{lemma}

In the statement above, for a finite tree $T_0$ with two marked, adjacent vertices $u$ and $v$, $N((u,v)) = T_0$ means that there is a graph isomorphism from $N((u,v))$ to $T_0$ that fixes the special vertices $u$ and $v$.

\begin{proof}  
 Let us build the connected component $N((u,v))$ by revealing gradually the edges appearing in $R(x)$, starting from the edges incident to $u$ and $v$. At any point, we take some vertex $w \in N((u,v))$ that we have not processed so far, and generate its incident edges to other vertices not processed so far. %If the neighborhood at any point is not consistent with the respective neighborhood in  $T_0$, we fail. 
 We couple this process with the tree process $\TT$, where we also start from the edge $(u,v)$ and iteratively add new children to a vertex not processed so far; in this process, the number of children of each node is a Poisson random variable with expectation $1$. As long as no cycle is created in $R(x)$, and the number of neighbors of $w$ is the same in $R(x)$ and $\TT$, we also fix some bijection between the children in both models, and hence maintain an isomorphism between the trees generated so far. We say in this case that the two processes are consistent; otherwise that the two processes deviate from each other, and the coupling fails.

 %We plan to couple this process we have just described with the process used to generate $\TT$, assuming $\TT$ is a tree isomorphic to $T_0$.
  %If the neighborhood at any point is not consistent with the respective neighborhood in  $T_0$, we fail. 

Suppose we are processing a vertex $w$, and the trees produced so far in $R(x)$ and $\TT$ are isomorphic.
With a slight abuse of notation, we denote by $w$ the respective vertex in each model, and 
by $T_w$ the tree produced so far in both models.
We generate the new incident edges to $w$ in each model and compare the two distributions. 

We prove that the probability that there exists a $w$ such that the two processes deviate from each other while processing vertex $w$ and $|T_w| \leq |T_0|$ is at most $3 \epsilon |T_0|^2$. 
 
This will prove that the probability that the two processes generate a particular tree $T_0$ can differ by at most $3 \epsilon |T_0|^2$. 

First, consider the event that there is an edge in $R(x)$ from $w$ back to another vertex in $T_w$ which creates a cycle. (Clearly, this cannot happen in the process $\TT$.) 
Since the probability of each edge is bounded by $\epsilon$, and the tree generated so far has at most $|T_0|$ vertices, the probability that $w$ has any incident edge in $R(x)$ that creates a cycle is at most $\epsilon |T_0|$. 
If this happens, the coupling fails.

The edges in $R(x)$ from $w$ to vertices not processed so far appear independently, and their expected number is $\sum_{w' \notin T_w: (w,w') \in E} x_{ww'}$. (In the following, we drop the condition $(w,w') \in E$ and assume that that $x_{ww'} = 0$ for any non-edge $(w,w')$.) 
Since $|T_w| \leq |T_0|$,  $x_{ww'} \leq \epsilon$ and $\sum_{w'} x_{ww'} = 1$, we have $\sum_{w' \notin T_w} x_{ww'} \in [1 - \epsilon |T_0|, 1]$. Thus the distribution of the number of new neighbors of $w$ (outside of $T_w$) is a summation of independent 0/1 Bernoulli variables with expectations bounded by $\epsilon$, and the total expectation is between $1 - \epsilon |T_0|$ and $1$. We can add one more Bernoulli variable with expectation at most $\epsilon |T_0|$, to obtain a summation of Bernoulli variables of total expectation $1$. Clearly, this affects the distribution by total variation distance at most $\epsilon |T_0|$. 
Then, by Le Cam's theorem, a summation of 0/1 Bernoulli variables of total expectation $1$ is close to a Poisson distribution with expectation $1$, within total variation distance $2 \sum_{w' \notin T_w} x_{ww'}^2 \leq 2 \epsilon$. Therefore, the number of children is within total variation distance $\epsilon |T_0| + 2\epsilon \leq  2\epsilon |T_0|$ of a Poisson variable of expectation $1$. (Recall that $|T_0| \geq 2$ because it always contains at least the two vertices $u,v$.) 

In other words, the probability that the number of children of $w$ in $R(x)$ differs from that in $\TT$ is at most $2 \epsilon |T_0|$. Also, as we argued above, the probability that $w$ is incident to any edge creating a cycle in $T_w$ is at most $\epsilon |T_0|$. By the union bound, the probability that the neighborhood of $w$ in $R(x)$ is different from the neighborhood in $\TT$ is at most $3 \epsilon |T_0|$.

Hence the processes in $R(x)$ and in $\TT$ can be coupled so that at each step, the probability that the two processes deviate from each other is at most $3\epsilon |T_0|$. As long as the two processes are consistent, we can extend the isomorphism defined so far, and hence the two trees in $R(x)$ and $\TT$ are isomorphic. By induction, the probability that the two processes ever deviate in the first $|T_0|$ steps is at most 3$\epsilon |T_0|^2$. Hence the probability that $N((u,v)) = T_0$ can differ from the probability that $T=T_0$ by at most $3\epsilon |T_0|^2$. 
\end{proof}

Lemma~\ref{lemma:randomtree} is false if we do not assume that $\|x\|_\infty \leq \epsilon$. It is also important to note that this lemma is valid only for simple graphs. If this lemma was true for graphs with parallel edges, we would be able to reduce the problem for an arbitrary fractional solution $x$ to the case of $\|x\|_\infty \leq \epsilon$ by replacing each edge by $1/\epsilon$ parallel edges. Unfortunately, this reduction doesn't work with our current approach.

\subsection{The Karp-Sipser algorithm on trees}
\label{sec:Karp-Sipsertrees}

It is easy to prove by induction (using the fact that trees always have degree 1 vertices) that in an execution of the Karp-Sipser algorithm on a forest, an edge must eventually either disappear, or else, is added to the matching. Together with the fact that the Karp-Sipser algorithm does not make mistakes, this shows that the Karp-Sipser algorithm finds a maximum matching in a tree.

Given a tree, we would like to be able to analyze which vertices and edges end up in the matching the algorithm selects, independent of the random choices the algorithm makes. To that end, consider the following algorithm to label the vertices of a tree either $L$ or $W$ (which is similar to, but not exactly same as the scheme in \cite{KS81}):

\paragraph{L/W-labeling algorithm.}
{\em Root the tree at an arbitrary vertex $r$. Consider any vertex $v$ at maximum possible depth that has not been labeled yet. If $v$ has no L-child (including the case where it has no children at all), label it L. Else, label it W. Repeat as long as there is any unlabeled vertex.}

\

The following claims are true (regardless of the chosen root, and regardless of the random choices the algorithm makes):

\begin{lemma}
For the L/W-labeling algorithm described above:
\begin{enumerate}
    \item If an edge between a W parent and an L child disappears, it must be because the W vertex has been matched.
    \item Every W vertex is matched at some point.
    \item Every edge between two W vertices disappears.
\end{enumerate}
\end{lemma}

\begin{proof}
\iffalse
{\em Claim 1.} Suppose by way of contradiction that an edge between a W parent and L child disappears because the L vertex has been matched. Certainly, this does not happen in the first step of the execution of the algorithm. Consider the very first time it happens.

The L vertex must have been added to the matching through a W labeled child it has. This W vertex must have degree 1, and so the edge connecting it to an L child must have disappeared. This contradicts our assumption of the original edge being the first edge between a W parent and an L child that has disappeared because the L vertex was added to the matching. \hfill \qedsymbol
\fi

\textit{Claim 1:} Suppose for a contradiction that an edge between a W parent $x$ and an L child $y$ disappears because another edge incident to $y$ is added to the matching. This must be an edge between $y$ and its own child $w$, labeled W (otherwise $y$ would not be labeled L.) Consider the first time this happens:
%The L vertex must have been added to the matching due to an edge to a W child being added to the matching. 
The W child $w$ must have degree 1 at the time it is matched, but it must have had at least one L child $z$ originally, because it is labeled W. Therefore, the edge $(w,z)$ disappeared earlier in the process, because $z$ was matched.
This contradicts our assumption that $(x,y)$ was the first edge between a W parent and an L child that disappeared because the L vertex was matched. % \hfill \qedsymbol

\textit{Claim 2.} Every W vertex $x$ is labeled as such due to an edge connecting it to an L child $y$. Either the edge $(x,y)$ disappears, and the claim follows from claim 1, or $(x,y)$ is added to the matching, and so the claim holds in both cases. 
% \hfill \qedsymbol

\textit{Claim 3.} Suppose by way of contradiction that an edge between two W vertices $(x,y)$ is added to the matching. Consider the state of the graph just before $(x,y)$ is added to the matching. One of the W vertices, say $y$, must have degree 1 at this point, but originally it had an L child $z$. So the edge $(y,z)$ must have disappeared. Again by claim 1, the only way such an edge can disappear is that the W vertex $y$ was matched earlier, a contradiction.
\end{proof}

\subsection{The Karp-Sipser algorithm on random trees}
\label{sec:Karp-Sipserrandomtrees}

We can now calculate the probability with which the Karp-Sipser algorithm, when executed on the random tree $\TT$ starting from the edge $(u,v)$, includes $(u,v)$ in the matching. This analysis is essentially identical to \cite{KS81} but we include it here for completeness, in a somewhat simplified form.

\begin{lemma}
\label{lemma:Karp-Sipser}
The Karp-Sipser algorithm, executed on $\TT$ initiated with the edge $(u,v)$, includes $(u,v)$ in the matching with probability exactly $2(1-\lambda) - \lambda^2$, where $\lambda$ is the unique real solution of $\lambda = e^{-\lambda}$.
\end{lemma}

\begin{proof}
We label the trees rooted at $u$ and $v$ using the procedure described in the previous section (imagining the special edge connecting $u$ and $v$ does not exist, and we are just labelling two different rooted trees).

Let us first calculate the probability that $u$ is labeled L (which we denote $\lambda$):
\begin{align*}
    \lambda &= \Pr[u \text{ is labeled L}] \\
    &= \Pr[u \text{ has no children labeled L}] \\
    &= \sum_{k=0}^{\infty}\Pr[u \text{ has $k$ children and none of them are labeled L}] \\
    &= \sum_{k=0}^{\infty}\frac{e^{-1}}{k!}(1-\lambda)^k = e^{-1}\cdot e^{1-\lambda} = e^{-\lambda}.
\end{align*}
$\lambda$ is thus the unique real number that solves the equation $\lambda = e^{-\lambda}$. 

Second, let us calculate the probability that the edge between $u$ and $v$ is added to the matching, and $v$ is labeled L. Imagine now rooting the random tree $\TT$ at $u$. This does not change the label of any of the vertices except possibly $u$.
Assuming that $v$ is labeled L, $u$ is now labeled W.
This means that $u$ must end up in the matching. None of the edges connecting $u$ with any of its W children end up in the matching. All the edges connecting $u$ with any of its L children, and the special edge between $u$ and $v$ are completely symmetric from the standpoint of the execution of the Karp-Sipser algorithm. Therefore,
\begin{eqnarray*}
    &&\Pr[\text{$(u, v)$ is added to the matching, $v$ is labeled L}]\\
    &=& \sum_{k=0}^{\infty} \Pr[\text{$(u, v)$ is added to the matching, $v$ is labeled L, $u$ has $k$ L children}]\\
    &=& \sum_{k=0}^{\infty}\frac{\lambda}{k+1}\sum_{r=k}^{\infty}\binom{r}{k}\lambda^k(1-\lambda)^{r-k}\frac{e^{-1}}{r!} = \sum_{k=0}^{\infty}\frac{\lambda^{k+1}e^{-\lambda}}{(k+1)!} 
    = (e^{\lambda}-1)e^{-\lambda}
    = 1-\lambda.
\end{eqnarray*}
Third, note that if we initially labeled both $u$ and $v$ as L (in the two separate trees rooted at $u$ and $v$), then $(u, v)$ must end up in the matching. This is because if we imagine rooting the tree at $u$,  $u$ is labeled W, so ends up in the matching, but the only way this can happen is if $(u, v)$ ends up in the matching since all of its other children are labeled W.

Fourth, note that if we initially labeled both $u$ and $v$ as W, then $(u, v)$ cannot be in the matching. This is because if we imagine rooting the tree at $u$,  the labelling remains the same, and every edge between two $W$ vertices disappears.

Therefore, we can compute the probability that the special edge $(u, v)$ ends up in the matching selected by Karp-Sipser as the sum of the probabilities of the edge ending up in the matching and $u$ and $v$ are labeled (initially) (L,L), (L,W), or (W,L). As we argued above, the probability that the edge appears in the matching and $u$ is labeled $L$ is $1-\lambda$. Symmetrically, the same holds when $v$ is labeled $L$. The probability of the union of the two events is $2(1-\lambda)$ minus the probability that both $u,v$ are labeled $L$ (in the initial labeling; in which case the edge is certainly selected). The probability that each of $u, v$ is labeled $L$ is $\lambda$, and these events are independent, since the two trees are generated and labeled independently. Hence the probability that both $u$ and $v$ are labeled $L$ is $\lambda^2$.
%We conclude that the probability with which  $(u, v)$ ends up in the matching selected by Karp-Sipser is exactly $2(1-\lambda) - \lambda^2$. Formally, 
In conclusion,
\begin{eqnarray*}
    &&\Pr[\text{$(u, v)$ is added to the matching}]\\
    &=& \Pr[\text{$(u, v)$ is added to the matching, $u$ is labeled L}] \\ && +\Pr[\text{$(u, v)$ is added to the matching, $v$ is labeled L}] \\&& - \Pr[\text{$(u, v)$ is added to the matching, $u, v$ are labeled L}]\\
    % &=& 2(1-\lambda) - \Pr[\text{$u, v$ are labeled L}]\\
    % &=& 2(1-\lambda) - \Pr[\text{$u$ is labeled L}]\Pr[\text{$v$ is labeled L}]\\
    &=& 2(1-\lambda) - \lambda^2.\\
\end{eqnarray*}
\end{proof}

\subsection{Putting it all together}
Here we finish the proof of Theorem~\ref{thm:Karp-Sipser}, which in turns implies Theorem~\ref{thm:0.544}.
The proof follows from a careful application of Lemma~\ref{lemma:randomtree}, which establishes a close correspondence between the neighborhood of an edge in the Galton-Watson process and the neighborhood of an edge in $R(x)$. This correspondence can be exploited to relate the behavior of the Karp-Sipser algorithm in the two settings. 

Let us consider a fractional matching $x$ such that $ \|x\|_\infty \leq \epsilon$.
Let us fix an edge $(u,v) \in E$ and denote by $\TT$ the random tree process starting from $(u,v)$.
Let $M$ denote the matching selected by the Karp-Sipser algorithm on the random graph $R(x)$, and $M'$ denote the matching selected on the tree $T \sim \TT$. Let $F = \{T_1, \ldots,T_m\}$ be any finite set of finite trees with marked vertices $u,v$ (possible realizations of the tree process $\TT$). By the triangle inequality, we have
\begin{eqnarray*}
    && \Big|\Pr_{R(x)}[(u, v)\in M\mid(u,v) \in R(x)] - \Pr_{T \sim \TT}[(u, v) \in M'] \Big|\\
    & =  & \Big|\Pr[(u, v)\in M, N(u,v) \notin F \mid (u,v) \in R(x)] - \Pr[(u,v) \in M', T \notin F] \\
    && + \Pr[(u, v)\in M, N(u,v) \in F \mid (u,v) \in R(x)] - \Pr[(u, v) \in M', T \in F] \Big| \\
    &\leq & \Pr[(u,v) \in M, N((u,v))\notin F \mid (u,v) \in R(x)] + \Pr[(u,v) \in M', T \notin F] \\
    && + \left|\Pr[(u, v)\in M, N(u,v) \in F \mid (u,v) \in R(x)] - \Pr[(u, v) \in M', T \in F]\right| \\
    &\leq& \Pr[N((u,v))\notin F\mid(u,v) \in R(x)] + \Pr[T \notin F] \\
    && + \left|\Pr[(u, v)\in M, N(u,v) \in F \mid (u,v) \in R(x)] - \Pr[(u, v) \in M', T \in F]\right|.
\end{eqnarray*}
We replaced the first two probabilities by probabilities of obviously larger events.
Next, we subdivide the last two probabilities further by considering each of the possible trees $T_i \in F$:
\begin{eqnarray*}
    && \left|\Pr[(u, v)\in M, N(u,v) \in F \mid (u,v) \in R(x)] - \Pr[(u, v) \in M', T \in F]\right| \\
    &=& \Big| \sum_{i=1}^m \Pr[(u, v)\in M \mid N(u,v) = T_i,  (u,v) \in R(x)]\Pr[N(u,v) = T_i \mid  (u,v) \in R(x)]\\
    &&- \sum_{i=1}^m \Pr[(u, v) \in M' \mid T = T_i] \Pr[T = T_i] \Big| \\
    &=& \Big|\sum_{i = 1}^m \Pr[(u, v) \in M \mid N(u,v) = T_i] \, (\Pr[N(u,v) = T_i \mid  (u,v) \in R(x)]- \Pr[T = T_i]) \Big|.
\end{eqnarray*}
Here, we used the fact that conditioned on the random graph $R(x)$ being a fixed tree $T_i$, the probability of selecting $(u,v)$ is exactly as on the tree process conditioned to be the same tree $T_i$. Finally, we use Lemma~\ref{lemma:randomtree} to bound the difference between the last two probabilities:
$$|\Pr_{R(x)}[N((u, v))  = T_i \mid (u,v) \in R(x)] - \Pr_{T \sim \TT}[T = T_i]| \leq 3\epsilon |T_i|^2.$$
From here, we also obtain
$$|\Pr_{R(x)}[N((u, v)) \notin F \mid (u,v) \in R(x)] - \Pr_{T \sim \TT}[T \notin F]| \leq 3\epsilon \sum_{T_i \in F}|T_i|^2.$$
To summarize, we have
\begin{eqnarray*}    
    \Big|\Pr_{R(x)}[(u, v)\in M\mid(u,v) \in R(x)] - \Pr_{T \sim \TT}[(u, v) \in M'] \Big|
    &\leq & 2\Pr_{T \sim \TT}[T \notin F] + 6 \epsilon \sum_{T_i \in F}|T_i|^2
\end{eqnarray*}
Recall also that we established in Lemma~\ref{lemma:Karp-Sipser} that $\Pr_{T \sim \TT}[(u,v) \in M'] = 2(1-\lambda) - \lambda^2.$

Fix any $\delta>0$. Since we know that $\TT$ produces a finite tree with probability 1, there is a finite set of finite trees $F$ such that 
\[\Pr_{T \sim \TT}[T \notin F] < \frac{\delta}{3},\]
and we can also choose $\eps > 0$ such that 
$$ 6\epsilon \sum_{T_i \in F}|T_i|^2 < \frac{\delta}{3}.$$
From the computations above, this implies that
\[\left|\Pr[(u, v)\in M\mid(u,v) \in R(x)] - (2(1-\lambda) - \lambda^2)\right| < \delta. \]
In other words, there is function $f$ with $\lim_{\eps \to 0}f(\eps) = 0$ such that 
$$ \left|\Pr[(u, v)\in M\mid(u,v) \in R(x)] -(2(1-\lambda) - \lambda^2)\right|\leq f(\eps).$$
Note that $f$ is defined independent of the graph and the specific fractional solution $x$. This establishes the first part of Theorem \ref{thm:Karp-Sipser}. 

For the second part of the theorem, once again, fix a set of finite trees $F$. Then,
\begin{eqnarray*}
    &&\Pr[N((u, v)) \text{ contains a cycle}\mid(u,v) \in R(x)]\\
    &\leq& \Pr[N((u, v)) \notin F\mid(u,v) \in R(x)] \\
    &\leq& \Pr[T \notin F] +  6\epsilon \sum_{T_i \in F}|T_i|^2.
\end{eqnarray*}
Following the same computations as above, we conclude that
$$\Pr[N((u, v)) \text{ contains a cycle}\mid(u,v) \in R(x)] \leq f(\eps).$$

It follows by linearity of expectation that
\begin{eqnarray*}
&&\mathbb E[\text{\# of edges in $R(x)$ whose connected components contain cycles}] \\
&=& \sum_{(u,v)}\Pr[N((u, v)) \text{ contains a cycle}\mid(u,v) \in R(x)]x_{uv} \\
&\leq& \sum_{(u,v)} f(\eps)x_{uv} \\
&=& \frac{f(\eps)|V(x)|}{2}.
\end{eqnarray*}
Now, since we know the Karp-Sipser algorithm finds a maximum matching in a tree, it follows that 
\begin{eqnarray*}
   &&\mathbb E[\text{size of maximum matching in } R(x)]\\
&\leq &\mathbb E[\text{size of matching selected by Karp-Sipser in } R(x)] +\frac{f(\eps)|V(x)|}{2}  \\
&\leq & \frac{((2(1-\lambda) - \lambda^2) +f(\eps))|V(x)|}{2} + \frac{f(\eps)|V(x)|}{2},
\end{eqnarray*}
which is exactly the desired result.
% Note that the same calculation works for the absolute value of the quantity considered. Since $F$ was arbitrary, and as we argued earlier, for large enough $F$, we can make $\Pr[T \in F]$ arbitrarily close to $1$, it follows that the left-hand side must be $0$, which is what we wanted to show.
% Therefore, the conditional probability of $(u,v)$ being selected in the matching is asymptotically (as $\epsilon \to 0$) the same in $R(x)$ as in the Galton-Watson tree, which we determined to be $2(1-\lambda) - \lambda^2$. 

\section{Improved contention resolution schemes for bipartite matchings}

Now we turn to contention resolution for bipartite matchings, without any assumption on the $\ell_\infty$ norm of the fractional matching.

\subsection{Contention resolution for $1$ element}

\label{sec:CRS-1} A basic building block of our CRSs is the following theorem which establishes the existence of a scheme for choosing $1$ out of $n$ elements (historically the first CRS \cite{Feige09}). 
\begin{theorem}
\label{theorem:CRS-1}
Suppose that $\cD$ is a distribution on $2^E$ such that for every set $S \subseteq E$, 
$$\Pr_{R \sim \cD}[S \cap R \neq \emptyset] \geq \sum_{i \in S} \beta_i.$$
Then there is a monotone contention resolution scheme for choosing one element $e(R)$ from $R \sim \cD$ such that $\Pr[e(R) = i] \geq \beta_i$  for every $i \in E$.
\end{theorem}

Recall that a CRS in this context is called monotone, if for every element $i \in E$,  $\Pr[e(R) = i]$ is non-increasing as a function on the sets $R$ containing $i$. Note that the theorem implies that unlike for CRSs for matchings, any CRS for selecting for choosing one element can be assumed to be monotone. This theorem can be proved using a max-flow/min-cut argument, as briefly discussed in \cite{Feige09}:
We set up a bipartite graph where the left-hand side is $2^E$ and the right-hand side is $E$.
We insert a directed edge for every pair $(S,i)$ where $i \in S$, with unbounded capacity. Finally, we add a source vertex $s$ which is connected to each vertex $S \in 2^E$ by an edge $(s,S)$ with a capacity of $\Pr_\cD[S]$, and a sink vertex $t$ which is connected to each vertex $i \in E$ by an edge $(i,t)$ of capacity $\beta_i$. By our assumption, it can be verified that there is no $s$-$t$ cut of capacity less than $\sum_{i \in E} \beta_i$. Hence, there exists a flow from $s$ to $t$ of capacity $\sum_{i \in E} \beta_i$, which can be interpreted as the desired contention resolution scheme by considering the flow through each vertex $S \in 2^E$.
\begin{figure}[ht] 
\centering
\resizebox{0.5\textwidth}{!}{%
\begin{tikzpicture}
    \tikzset{node style/.style={draw, circle, minimum size=0.8cm}}
\small
    % Source vertex
    \node[draw, circle] (s) at (0,0) {$s$};

    % Vertices in 2^E with uniform size
    \node[node style] (empty) at (2,2.2) {$\emptyset$};
    \node[node style] (U) at (2,1) {$\{u\}$};
    \node[node style] (V) at (2,-1) {$\{v\}$};
    \node[node style] (UV) at (2,-2.2) {$\{u,v\}$};
    % Vertices in E
    \node[draw, circle] (u) at (5,1) {$u$};
    \node[draw, circle] (v) at (5,-1) {$v$};

    % Sink vertex
    \node[draw, circle] (t) at (7,0) {$t$};

  % Edges from source to 2^E vertices
    \draw[->] (s) -- node[pos=0.7, left] {$\Pr_\cD[\emptyset]$} (empty);
    \draw[->] (s) -- node[pos=0.2, right] {\ \ $\Pr_\cD[\{u\}]$} (U);
    \draw[->] (s) -- node[pos=0.2, right] {\ \ $\Pr_\cD[\{v\}]$} (V);
    \draw[->] (s) -- node[pos=0.7, left] {$\Pr_\cD[\{u,v\}]$} (UV);

    % Edges from 2^E vertices to E vertices
    \draw[->] (U) -- node[above] {} (u);
    \draw[->] (V) -- node[above] {} (v);
    \draw[->] (UV) -- node[above left] {} (u);
    \draw[->] (UV) -- node[above right] {} (v);

    % Edges from E vertices to sink
    \draw[->] (u) -- node[above] {$\beta_u$} (t);
    \draw[->] (v) -- node[above] {$\beta_v$} (t);
\end{tikzpicture}
}

\caption{Bipartite graph to establish Theorem \ref{theorem:CRS-1} with $|E| = 2$.}
\label{fig:theorem4}
\end{figure}
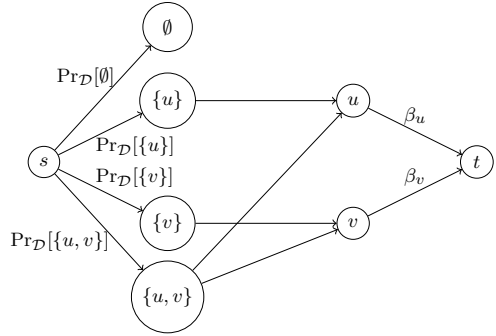

We remark that the resulting scheme can be assumed to be monotone: Denote $\pi_i(S) = \Pr[e(R)=i \mid R=S]$, and consider a scheme as above minimizing the potential function $$\Phi = \sum_{i=1}^{n} \sum_{S_1 \subset S_2, i \in S_1} \Pr[R=S_1] \Pr[R=S_2] (\pi_i(S_2) - \pi_i(S_1))_+.$$ For any two sets $S_1 \subset S_2, |S_2| = |S_1|+1$, if $i \in S_1$ such that $\pi_{i}(S_1) < \pi_{i}(S_2)$, there is another element $i' \in S_1$ such that $\pi_{i'}(S_1) > \pi_{i'}(S_2)$, and hence we can modify the scheme by transferring some survival probability from $\pi_{i'}(S_1)$ to $\pi_{i}(S_1)$ and from $\pi_{i}(S_2)$ to $\pi_{i'}(S_2)$, thereby reducing the potential function $\Phi$.

\subsection{A $0.480$-balanced scheme for bipartite matchings}

Before going to the proof of Theorem~\ref{thm:0.509}, we will show the existence of a simple $2(1-e^{-1/e}) - e^{-2}$-balanced contention resolution scheme for bipartite matchings\footnote{Note the similarity in the expression to $2(1-\lambda)-\lambda^2 = 2(1-e^{-\lambda})-\lambda^2$, the constant from our previous analysis. This similarity is not a coincidence, and we can think of the scheme we describe as first-order approximation to Karp-Sipser.}. We remark that $2(1-e^{-1/e}) - e^{-2} \geq 0.480$ and hence this already beats the {\em optimal monotone scheme} for bipartite matchings. By necessity, our scheme is non-monotone and this result establishes a strict separation between monotone and non-monotone schemes for bipartite matchings.
\paragraph{The simple scheme:}
\begin{enumerate}
    \item For each edge $(u, v)$ (with probability $x_{uv}$ of appearing), independently declare it active with probability $\frac{1-e^{-x_{uv}}}{x_{uv}}$ given it appears.
    \item For each vertex $u$, call an active edge $(u, v)$ ``available at vertex $u$'' if $v$ has no other edges adjacent to it which are active.
    \item Using a contention resolution scheme for $1$ element, select one of the available edges at each vertex, if any. (We claim and prove that this can be done so that $(u, v)$ gets selected at vertex $u$ with probability at least $x_{uv} \left((1-e^{-1/e})-\frac{1}{2e^2}\right) +\frac{e^{2x_{uv}}-e^{x_{uv}}}{2e^2}$.)
    \item The set of edges selected at all the vertices form the matching $M$.
\end{enumerate}

\

The first step of the scheme is a pre-processing step which ensures that edges with high probabilities of appearing don't destroy the chances of neighboring edges getting picked by the scheme. This strategy has also been used in the literature previously \cite{BZ22}. The second step is a first attempt at using the idea that it is always useful to add vertices with degree 1 into the matching. This does not seem to have been explicitly exploited by previous CRSs.

To analyze the algorithm, note firstly that the selected edges form a matching $M$:  If $M$ contained two edges incident to a vertex $u$, then only one of them could have been selected as available at vertex $u$ (by nature of the contention resolution scheme at $u$). Therefore, the other edge, say $(u,w)$, was selected as available at vertex $w$. But this is a contradiction, because if $(u,w)$ is available at $w$, there cannot be any other active edge incident to $u$.

Secondly, let us note that the probability that an edge $(u,v)$ is available at a vertex $u$ is
$$F(x_{uv}) = \left(1-e^{-x_{uv}}\right)\prod_{u' \in \delta (v), u' \neq u} e^{-x_{u'v}}=\left(1-e^{-x_{uv}}\right) e^{-(1-x_{uv})} =  \frac{e^{x_{uv}}-1}{e}$$
and similarly, the probability that an edge $(u,v)$ is isolated amongst active edges is $$\left(1-e^{-x_{uv}}\right)\left( e^{-(1-x_{uv})}\right)^2 = \frac{e^{2x_{uv}}-e^{x_{uv}}}{e^2}$$

Hence, if a CRS of the sort claimed in step 3 exists, the desired result follows since
\begin{eqnarray*}
    &&\Pr[(u,v) \in M]\\ &=& 2\Pr[(u,v) \text{ is selected at } u] - \Pr[(u,v) \text{ is selected at both } u, v] \\
    & =& 2\Pr[(u,v) \text{ is selected at } u] - \Pr[(u,v) \text{ is isolated amongst active edges}] \\
    &\geq& 2x_{uv} \left((1-e^{-1/e})-\frac{1}{2e^2}\right) +\frac{e^{2x_{uv}}-e^{x_{uv}}}{e^2}-\frac{e^{2x_{uv}}-e^{x_{uv}}}{e^2}\\
    &=& 2x_{uv}\left((1-e^{-1/e})-\frac{1}{2e^2}\right)
\end{eqnarray*}
Furthermore whether $(u, v_i)$ is available at $u$ is independent of whether $(u, v_j)$ is available at $u$ (this is where we use the fact that that the graph is bipartite; note that we actually only need to assume that the graph is triangle-free; it is unclear how to drop this assumption). Therefore, it follows from Theorem~\ref{theorem:CRS-1} that if the probability of $(u, v_i)$ appearing is $x_i$ (short for $x_{uv_i}$), the existence of the required CRS depends only on whether 
\begin{align*}
    \Pr[\text{at least one of } (u, v_i) \text{ is available}] &= 1-\prod_{i=1}^n(1-F(x_i))\\& \geq \sum_{i = 1}^nx_i\left((1-e^{-1/e})-\frac{1}{2e^2}\right) +\frac{e^{2x_i}-e^{x_i}}{2e^2}.
\end{align*}
Therefore, the desired result depends on showing that the infimum of 
\[\left(1-\prod_{i=1}^n(1-F(x_i)) - \sum_{i = 1}^n \frac{e^{2x_i}-e^{x_i}}{2e^2}\right) \cdot \frac{1}{\sum_{i=1}^n x_i}\]
for $x_i \geq 0$ and $\sum_{i=1}^n x_i \leq 1$ is $1-e^{-1/e}-\frac{1}{2e^2}$, attained when all the $x_i$ are small and sum to $1$. This follows from the following lemma.

\begin{lemma}
\label{lem:bound}
If $\sum x_i \leq 1, x_i \geq 0, $ and $F(x) =\frac{e^x-1}{e}$, then
$$1-\prod_{i=1}^n(1-F(x_i)) \geq \sum_{i=1}^n x_i\left(1-e^{-1/e}-\frac{1}{2e^2}\right) + \sum_{i=1}^{n} \frac{e^{2x_i}-e^{x_i}}{2e^2}.$$
\end{lemma}

\begin{proof}
Let us assume without loss of generality that $0 \leq x_1 \leq x_2 \leq \cdots \leq x_n$,  and let us write $A_i = e^{-x_i/e}, B_i = 1-F(x_i)$. Now we know that $$e^{-\sum x_i/e} -\prod_{i=1}^n(1-F(x_i)) = \prod A_i - \prod B_i = \sum_{j=1}^n \left(\frac{A_j - B_j}{B_j} \prod_{i \leq j}B_i \prod_{i>j}A_i\right)$$

\begin{claim}
Assuming $0 \leq x_1 \leq x_2 \leq \cdots \leq x_n$ and $\sum x_i \leq 1$, we have that 
$$\prod_{i \leq j}B_i \prod_{i>j}A_i \geq B_j^{1/x_j}.$$
\end{claim} 

\begin{proof} Let us write $S = \sum_{i \leq j}x_i$. Then we know $$\log \left(\prod_{i>j}A_i\right) 
= -\sum_{i>j} \frac{x_i}{e}
= -\frac{\sum x_i-S}{e}\geq -\frac{1-S}{e}$$

Furthermore, since  $\frac{\log (1-F(x))}{x}$ is a decreasing function, $\frac{\log B_i}{x_i} \geq \frac{\log B_j}{x_j}$. It follows that 
$$\log \left(\prod_{i\leq j}B_i\right)= \sum_{i \leq j} \log B_i \geq \frac{\log B_j}{x_j} \sum_{i \leq j} x_i 
= \frac{\log B_j}{x_j}S$$

Finally, since $-\frac{1}{e}\geq \frac{\log B_j}{x_j}$ (once again because $\frac{\log (1-F(x))}{x}$ is a decreasing function) it follows that 
$$ \log \left( \prod_{i \leq j}B_i \prod_{i>j}A_i \right) 
\geq \frac{\log B_j}{x_j}S -  \frac{1-S}{e} \geq \frac{\log B_j}{x_j}$$ 
\end{proof}

To finish the proof of Lemma~\ref{lem:bound}, we can write 
$$e^{-\sum x_i/e} -\prod_{i=1}^n(1-F(x_i)) \geq  \sum_{j=1}^n \frac{A_j - B_j}{B_j} B_j^{1/x_j}.$$
One can verify that $\frac{A_j - B_j}{B_j} B_j^{1/x_j} \geq \frac{e^{2x_j}-e^{x_j}-x_j}{2e^2}$, since the difference of the two quantities is an increasing function of the single variable $x_j \in [0,1]$.
Using this fact, we obtain
$$e^{-\sum x_i/e} -\prod_{i=1}^n(1-F(x_i)) \geq  \sum_{j=1}^n \frac{e^{2x_j}-e^{x_j}-x_j}{2e^2}.$$
We also know that since $0 \leq \sum x_i \leq 1$,  by convexity we have 
$$1-e^{-\sum x_i/e} \geq (1-e^{-1/e}) \sum x_j.$$
Adding up the last two inequalities, we obtain
$$ 1 - \prod_{i=1}^n(1-F(x_i)) \geq \left(1 - e^{-1/e} - \frac{1}{2e^{2}} \right)  \sum_{j=1}^{n} x_j
+ \sum_{j=1}^{n} \frac{e^{2x_j} - e^{x_j}}{2e^2}.$$
\end{proof}

There are many ways to improve this scheme. For example, the scheme completely ignored edges all of whose neighbors are edges associated with vertices of degree at least 2. We could use some other contention resolution scheme, for example, the one from \cite{BZ22} to add some of these edges to the matching generated by our scheme. However, we do not investigate this further here. Instead, we will consider a different approach.

\subsection{A $0.509$-balanced scheme for bipartite matchings}

In this section we present our best scheme for bipartite matchings in general. This beats the best possible random order online contention resolution scheme. 

Let us call the set of vertices on the left $V_1$, and on the right, $V_2$. There are two main ideas. The first idea is to select edges in two stages, with the first stage devoted to running contention resolution on the edges at each vertex in $V_1$, and the second stage is devoted to running contention resolution on the edges picked in stage 1 at each vertex in $V_2$. The other main idea involves noticing that edges $(u,v)$ such that $v$ has degree $1$ in $R$ have no competition in the second stage and hence should be preferentially selected in the first stage, since if selected they will certainly survive in our matching. 

\subsubsection{A warm-up: the 2-stage scheme}
We start by describe a simple scheme which achieves a factor of $\alpha = 1 - 1/e^{1-1/e} \simeq 0.468$ (due to Chandra Chekuri~\cite{Chekuri22}) illustrating the first idea.

\paragraph{The 2-stage scheme.}
\begin{enumerate}
\item For each vertex $u \in V_1$ independently, perform contention resolution among the edges $(u,v) \in R$, to choose one edge incident to $u$; call these edges $R_1$.
\item For each vertex $v \in V_2$ independently, perform contention resolution among the edges $(u,v) \in R_1$, to choose one edge incident to $v$; call these edges $M$.
\end{enumerate}

It is easy to see that $M$ is a matching, since the two stages of contention resolution ensure that the degrees in $M$ are at most $1$ on both sides.

\begin{lemma}
The first stage of contention resolution can be implemented so that for each edge, $$\Pr[(u,v) \in R_1] \geq (1-1/e) x_{uv}.$$ Also, these events are mutually independent for a fixed value of $v$.
\end{lemma}

\begin{proof}
For any subset of edges $S$ incident to $u$, $$\Pr[S \cap R' \neq \emptyset] = 1 - \prod_{(u,v) \in S} (1-x_{uv}) \geq 1 - e^{-\sum_{(u,v) \in S} x_{uv}} \geq (1-1/e) \sum_{(u,v) \in S} x_{uv},$$ by the concavity of the function $1 - e^{-t}$. By Theorem~\ref{theorem:CRS-1}, there is a scheme for each fixed $u \in V_1$ such that every edge $(u,v)$ survives with probability at least $(1-1/e) x_{uv}$. We do this independently for each $u \in V_1$, so there is no correlation between different edges $(u,v)$ for a fixed $v$.
\end{proof}

\begin{lemma}
The second stage of contention resolution can be implemented so that for each edge, $$\Pr[(u,v) \in M] \geq (1-1/e^{1-1/e}) x_{uv}.$$
\end{lemma}

\begin{proof}
Consider a fixed vertex $v \in V_2$. 
We have $\Pr[(u,v) \in R_1] \geq (1-1/e) x_{uv}$, and these events are independent for different edges incident to $v$. Hence, for any subset $S$ of such edges, 
$$\Pr[S \cap R_1 \neq \emptyset] \geq 1 - \prod_{(u,v) \in S} (1 - (1-1/e) x_{uv}) \geq 1 - e^{-(1-1/e) \sum_{(u,v) \in S} x_{uv}}.$$
Again by the concavity of the function $1 - e^{-(1-1/e) t}$, we have $\Pr[S \cap R_1 \neq \emptyset] \geq (1 - e^{-(1-1/e)}) \sum_{(u,v) \in S} x_{uv}$. Hence by Theorem~\ref{theorem:CRS-1}, there is a scheme such that each edge incident to $v$ survives with probability $\Pr[(u,v) \in M] \geq (1 - e^{-(1-1/e)}) x_{uv}$.
\end{proof}

\subsubsection{Improving the 2-stage scheme}

Next, we design a more complicated scheme which builds upon the 2-stage scheme. Similar to our first scheme, we start by letting an edge be``active'' with probability $1 - e^{-x_{uv}}$.  We mark the edges which are not active {\em gray}. Then, we mark active edges with degree $1$ on the right-hand side {\em red} with a certain probability, so that $\Pr[(u,v) \mbox{ is red}] = 1 - e^{-x_{uv}/e}$. This is somewhat lower than the probability of having degree $1$ on the right-hand side, but just enough to guarantee a probability of survival $(1-e^{-1/e}) x_{uv}$ among red edges, which is the optimal factor. We mark the remaining active edges blue and run a contention resolution scheme amongst them separately. At the end, all surviving red, blue and gray edges enter a contention resolution scheme at each vertex on the right-hand side.

\paragraph{The red/blue/gray scheme:}
\begin{enumerate}
\item Recall that $\Pr[(u,v) \in R] = x_{uv}$. Decide for each edge $(u,v) \in R$ independently at random whether to mark it {\bf gray}, so that $\Pr[(u,v) \mbox{ is gray}] = x_{uv} - (1-e^{-x_{uv}})$. We call the edges $(u,v) \in R$ which are not gray {\em active}.

\item For each $(u,v)$ such that $(u,v)$ is the only active edge incident to $v$, decide independently at random whether to mark $(u,v)$ {\bf red}, so that $\Pr[(u,v) \mbox{ is red}] = 1 - e^{-{x_{uv}/e}}$. Mark all other active edges {\bf blue}. We have $\Pr[(u,v) \mbox{ is blue}] = e^{-x_{uv}/e} - e^{-x_{uv}}$.

\item For each $u \in V_1$, if there are any red edges incident to $u$, perform contention resolution to include one of them in $R_1$, so that $\Pr[(u,v) \in R_1] = (1-e^{-1/e}) x_{uv}$.

\item For each $u \in V_1$, if there are no red edges incident to $u$, and there are some blue edges incident to $u$, perform contention resolution to include one of them in $R_2$, so that $\Pr[(u,v) \in R_2] = (e^{-1/e} - e^{-1}) x_{uv}$. 

\item For each $u \in V_1$, if there are no active (red or blue) edges incident to $u$, and there are some gray edges incident to $u$, perform contention resolution to include one of them in $R_3$, so that $\Pr[(u,v) \in R_3] \geq \frac{1}{2e} x_{uv}^2$.

\item Finally, for each $v \in V_2$, perform contention resolution among all edges in $R_1 \cup R_2 \cup R_3$ incident to $v$, to include one of them in $M$, so that $$ \Pr[(u,v) \in M] \geq 0.509 \ x_{uv}.$$
\end{enumerate}

Implicit in step 2 is the claim that the definition of red edges is valid. Implicit in each of steps 3, 4, 5, and 6 above is a claim that there exists a certain contention resolution scheme for choosing 1 out of $n$ elements. The existence of such schemes can be proved by applying Theorem~\ref{theorem:CRS-1}, if we can calculate the probability that at least one of a subset of edges ``is available for consideration at that stage" (i.e., is red, is blue, etc).

For steps 3, 4, and 5, this quantity is fairly simple to calculate, because all the choices (to designate edges red or blue or gray) are made independently. The following lemmas are devoted to establishing the claimed existences in these cases.

\begin{lemma}
The definition of red edges is valid and there is a CRS among red edges such that $\Pr[(u,v) \in R_1] = (1-e^{-1/e}) x_{uv}$ for every edge $(u,v)$. 
\end{lemma}

\begin{proof}
The probability that $(u,v)$ is the only active edge incident to $v$ is $$(1 - e^{-x_{uv}}) \prod_{u' \in V_1 \setminus \{u\}} e^{-x_{u'v}} = (e^{x_{uv}} - 1) \prod_{u' \in V_1} e^{-x_{u'v}} = (e^{x_{uv}} - 1) e^{-1}.$$ By elementary inequalities, this is at least $x_{uv} e^{-1} \geq 1 - e^{-x_{uv} / e}$. Therefore, the notion of red edges is well defined.

For every subset of edges $S$ incident to $u$, we have 
\begin{align*}
    \Pr[S \mbox{ contains a red edge}] &= 1 - \prod_{(u,v) \in S} e^{-x_{uv} / e} \\
    &= 1 - e^{-\sum_{(u,v) \in S} x_{uv} / e} \\
    &\geq (1 - e^{-1/e}) \sum_{(u,v) \in S} x_{uv}.
\end{align*}

 By Theorem~\ref{theorem:CRS-1}, there is a CRS among the red edges incident to $u$ such that each edge survives as a red edge with probability $(1 - e^{-1/e}) x_{uv}$. 
\end{proof}

\begin{lemma}
There is a CRS among blue edges such that 
$\Pr[(u,v) \in R_2] = (e^{-1/e} - e^{-1}) x_{uv}$ for every edge $(u,v)$. 
\end{lemma}

\begin{proof}
Each edge is active with probability $1 - e^{-x_{uv}}$ and marked red with probability $1 - e^{-x_{uv} /e}$; otherwise it is blue, hence $\Pr[(u,v) \mbox{ is blue}] = e^{-x_{uv} / e} - e^{-x_{uv}}$. 

Consider a set of edges $S$ incident to a vertex $u \in V_1$ and condition on the event that there are no red edges incident to $u$. Since the events of different edges incident to $u$ being red or blue are independent,
\begin{eqnarray*}
    &&\Pr[(u,v) \mbox{ is blue} \mid \mbox{no red edges incident to } u] \\ 
    &=& \Pr[(u,v) \mbox{ is blue} \mid (u,v) \mbox{ is not red}] \\
    &=& \frac{e^{-x_{uv} / e} - e^{-x_{uv}}}{e^{-x_{uv} / e}}\\
    &=& 1 - e^{-(1-1/e) x_{uv}}.
\end{eqnarray*}

This happens independently for each edge incident to $u$ (since $(u,v)$ being red/blue depends only on edges incident to $v$), and so we obtain
\begin{eqnarray*}
    && \Pr[\mbox{there is a blue edge among } S \mid \mbox{no red edges incident to } u] \\
    &=&1 - \prod_{(u,v) \in S} \Pr[(u,v) \mbox{ is not blue} \mid \mbox{no red edges incident to }u ]\\
    &=&  1 - e^{-(1-1/e) \sum_{(u,v) \in S} x_{uv}}\\
    &\geq& (1 - e^{-(1-1/e)}) \sum_{(u,v) \in S} x_{uv}.
\end{eqnarray*}

Therefore, by Theorem~\ref{theorem:CRS-1} there is a CRS among blue edges which, conditioned on no red edges being incident to $u$, includes each edge in $R_2$ with probability $(1 - e^{-(1-1/e)}) x_{uv}$. The probability of no red edge being incident to $u$ is $\prod_{v \in V_2} e^{-x_{uv} / e} = e^{-1/e}$. Therefore, 
$$\Pr[(u,v) \in R_2] = e^{-1/e} (1 - e^{-(1-1/e)}) x_{uv} = (e^{-1/e} - e^{-1}) x_{uv}.$$
\end{proof}

\begin{corollary}
\label{cor: beta}
$\Pr[(u,v) \in R_2 \mid (u,v) \mbox{ is blue}] \geq \beta = \frac{e^{-1/e}-e^{-1}}{1-e^{-1}}$.
\end{corollary}

\begin{proof}
We have $\Pr[(u,v) \mbox{ is blue}] = e^{-x_{uv}/e} - e^{-x_{uv}}$, and $\Pr[(u,v) \in R_2] = (e^{-1/e} - e^{-1}) x_{uv}$.
Recall that only blue edges can appear in $R_2$. Hence,
\begin{align*}
     \Pr[(u,v) \in R_2 \mid (u,v) \mbox{ is blue}] &= \frac{(e^{-1/e} - e^{-1}) x_{uv}}{e^{-x_{uv}/e} - e^{-x_{uv}}}\\
 &\geq \frac{(e^{-1/e} - e^{-1}) x_{uv}}{1 - e^{-(1-1/e)x_{uv}}}\\ &\geq 
 \frac{e^{-1/e} - e^{-1}}{1-e^{-1}}. 
\end{align*}

\end{proof}

\begin{lemma}
There is a CRS among gray edges such that
$\Pr[(u,v) \in R_3] \geq \frac{1}{2e} x_{uv}^2$ for every edge $(u,v)$. 
\end{lemma}

\begin{proof}
Consider a fixed $u \in V_1$ and let's condition on no active edges being incident to $u$; this happens with probability $\prod_{v \in V_2} e^{-x_{uv}} = e^{-1}$. Conditioned on this, $(u,v)$ is not a gray edge if and only if $(u,v) \notin R$. Therefore, for a set of edges $S$ incident to $u$,
\begin{eqnarray*}
&&\Pr[\mbox{no gray edge among } S \mid \mbox{no active edge incident to }u] \\
&=&\Pr[S \cap R = \emptyset \mid \mbox{no active edge incident to }u]\\
&=&\frac{\prod_{(u,v) \in S} (1-x_{uv})}{\prod_{(u,v) \in S} e^{-x_{uv}}}
 = e^{\sum_{(u,v) \in S} (x_{uv} + \ln (1-x_{uv}))}.
\end{eqnarray*}
Consider the Taylor expansion of the exponent:
$$ x_{uv} + \ln (1-x_{uv}) = -\sum_{k=2}^{\infty} \frac{x_{uv}^k}{k}.$$
We observe that this is a concave function, not only as a function of $x_{uv}$ but even as a function of $x_{uv}^2$. 
Therefore, if we fix $\sum_{(u,v) \in S} x_{uv}^2 = \sigma^2$, the maximum value that the exponent could achieve is attained for $x_{uv} = \frac{1}{\sqrt{|S|}} \sigma$ for all $(u,v) \in S$, in which case
$$ \sum_{(u,v) \in S} (x_{uv} + \ln (1-x_{uv})) = \sqrt{|S|} \sigma + |S| \ln (1 - \frac{\sigma}{\sqrt{|S|}}).$$
Again, we want to determine the maximum possible value of the right-hand side for given $\sigma$.
Observe that since $\sum_{(u,v) \in S} x_{uv} = \sigma \sqrt{|S|} \leq 1$, we have $|S| \leq 1 / \sigma^2$.
The right-hand side is an increasing function of $|S|$, hence it is maximized when $|S| = 1 / \sigma^2$ (possibly not an integer, but certainly this gives an upper bound):
$$ \sum_{(u,v) \in S} (x_{uv} + \ln (1-x_{uv})) \leq 1 + \frac{1}{\sigma^2} \ln (1 - \sigma^2).$$
Further, we can bound
$$ 1+ \frac{1}{\sigma^2} \ln (1 - \sigma^2) \leq \ln (1 - \frac12 \sigma^2) $$
for example by comparing the Taylor series. We conclude that
\begin{eqnarray*}
&&\Pr[\mbox{no gray edge among } S \mid \mbox{no active edge incident to }u] \\
&=& e^{\sum_{(u,v) \in S} (x_{uv} + \ln (1-x_{uv}))}\\
&\leq& 1 - \frac12 \sigma^2 \\
&=& 1 - \frac12 \sum_{(u,v) \in S} x_{uv}^2.
\end{eqnarray*}
By Theorem~\ref{theorem:CRS-1}, we obtain the desired CRS.
\end{proof}

To finish, we need to analyze Step 6 of the algorithm, which is contention resolution among all the surviving edges on the right-hand side. Here, there can be at most one red edge incident to a vertex $v \in V_2$, possibly multiple gray edges which appear independently, and possibly multiple blue edges whose survival up to this stage of the scheme is correlated. This correlation causes the main trouble in our analysis of this final step, because it makes it harder to calculate the probability that at least one of a subset of edges incident at a vertex $v$ is in $R_1\cup R_2 \cup R_3$. Ideally, we would like to prove that the appearance of blue edges satisfies some form of negative correlation. At the moment, we are able to prove only {\em pairwise} negative correlation which is sufficient to achieve the factor of $0.509$. A stronger correlation result (for example negative cylinder dependence) would lead to an improved factor.

\begin{lemma}
\label{lem:neg-correl}
For any two incident edges $(u,v)$ and $(u',v)$,
$$ \Pr[(u,v) \in R_2 \ \& \ (u',v) \in R_2 \mid (u,v),(u',v) \mbox{ are blue}] $$ is less than or equal to
$$\Pr[(u,v) \in R_2 \mid (u,v),(u',v) \mbox{ are blue}] \cdot \Pr[(u',v) \in R_2 \mid (u,v),(u',v)  \mbox{ are blue}].$$
\end{lemma}

\begin{proof}
Define $\Gamma(u) = \{ v': (u,v') \mbox{ active} \}$ and $\Gamma(u') = \{ v': (u',v') \mbox{ active} \}$. 
Note that conditioning on $(u,v), (u',v)$ being blue edges is the same as conditioning on $v \in \Gamma(u) \cap \Gamma(u')$, because edges $(u,v), (u',v)$ being active also means that they must be blue.

We claim that conditioned on $\Gamma(u), \Gamma(u')$ such that  $v \in \Gamma(u) \cap \Gamma(u')$, the probability that $(u,v) \in R_2$ is decreasing in $\Gamma(u)$ and increasing in $\Gamma(u')$, while conversely the probability that $(u',v) \in R_2$ is increasing in $\Gamma(u)$ and decreasing in $\Gamma(u')$. 

We prove this by considering a fixed choice of the active edges incident to $V_1 \setminus \{u,u'\}$, and at the end averaging over these choices. Consider $\Gamma(u), \Gamma(u')$ where $v \in \Gamma(u) \cap \Gamma(u')$.
For $(u,v)$ to be selected in $R_2$, there cannot be any red edge incident to $u$. The only candidates for such red edges are $(u,\tilde{v})$ where $\tilde{v} \in \Gamma(u) \setminus \Gamma(u')$, because edges incident to $\Gamma(u) \cap \Gamma(u')$ are blue by definition. For each $\tilde{v} \in \Gamma(u) \setminus \Gamma(u')$, $(u,\tilde{v})$ is red if $\tilde{v}$ does not have any other incident active edges (and an additional independent coin flip succeeds, as defined in Step 2). Clearly, the event of no red edge incident to $u$ is monotonically decreasing in $\Gamma(u) \setminus \Gamma(u')$.

In case there is no red edge incident to $u$, we perform contention resolution among the blue edges incident to $u$, which are all the edges $(u,v'), v' \in \Gamma(u)$. Since this scheme is monotone, the probability of survival is monotonically decreasing in $\Gamma(u)$.  This monotonicity property also remains preserved when we average over the choices of active edges incident to $V_1 \setminus \{u,u'\}$.
Overall, the probability of $(u,v)$ surviving in $R_2$ is monotonically decreasing in $\Gamma(u)$ and increasing in $\Gamma(u')$. Symmetrically, the probability of $(u',v)$ surviving in $R_2$ is monotonically decreasing in $\Gamma(u')$ and increasing in $\Gamma(u)$. 

Given this monotonicity property, we use the FKG inequality \cite{FKG} to prove our result\footnote{The FKG inequality states (in particular) that if $f$ and $g$ are two functions from $\{0,1\}^m$ to $\mathbb{R}$, $f$ increasing and $g$ decreasing on $\{0,1\}^m$, and $X$ is distributed according to a product distribution on $\{0,1\}^m$, then $\E[f(X)g(X)] \leq \E[f(X)]\E[g(X)]$.}. The appearances of vertices in $\Gamma(u)$ and $\Gamma(u')$ are independent both between $u,u'$ and for different vertices (since these are determined by the edges incident to $u$ and $u'$ respectively). Let us define $\gamma \in \{0,1\}^{2n}$ where $\gamma_i = 0$ iff $i \in \Gamma(u)$ and $\gamma_{i+n} = 1$ iff $i \in \Gamma(u')$; then $\gamma$ is distributed according to a product distribution. 
Also, conditioned on $\gamma$, i.e. a fixed choice of $\Gamma(u)$ and $\Gamma(u')$, the event $(u,v) \in R_2$ is independent of the event $(u',v) \in R_2$, because the contention resolution on edges incident to $u$ and $u'$ is performed independently. 
As we argued, $\Pr[(u,v) \in R_2 \mid \gamma]$ is increasing in $\gamma$ and $\Pr[(u',v) \in R_2 \mid \gamma]$ is decreasing in $\gamma$, for all $\gamma$ consistent with $v \in \Gamma(u) \cap \Gamma(u')$. 
We apply the FKG inequality to this subcube ($\gamma \in \{0,1\}^{2n}$ such that $\gamma_v = 0$ and $\gamma_{v+n} = 1$, i.e.~$v \in \Gamma(u) \cap \Gamma(u')$). We use $f(\gamma) = \Pr[(u,v) \in R_2 \mid \gamma]$ and $g(\gamma) = \Pr[(u',v) \in R_2 \mid \gamma]$:
\begin{eqnarray*}
 &&  \Pr[(u,v) \in R_2 \ \& \ (u',v) \in R_2 \mid v \in \Gamma(u) \cap \Gamma(u')] \\
 &=&  \E[f(\gamma) g(\gamma) \mid \gamma_v = 0, \gamma_{v+n} = 1] \\
& \leq& \E[f(\gamma) \mid \gamma_v = 0, \gamma_{v+n} = 1] \cdot \E[g(\gamma) \mid \gamma_v =0, \gamma_{v+n} = 1] \\
& =& \Pr[(u,v) \in R_2 \mid v \in \Gamma(u) \cap \Gamma(u')] \cdot \Pr[(u',v) \in R_2 \mid v \in \Gamma(u) \cap \Gamma(u')].
\end{eqnarray*}
as desired.\end{proof}

The main takeaway from this lemma is that if we let $\beta$ be a lower bound on the probability that an edge survives in $R_2$ given that it is blue, then, conditioned on having at least two active (and hence blue) edges at a vertex $v$ in $V_2$ in step 6, the probability that one of them survives in $R_2$ is at least $2\beta - \beta^2$. In the final analysis of step 6, if there are more than $2$ blue edges at a vertex $v$, we only use two of them. This allows us to establish our desired conclusion.

\begin{theorem}
There is a CRS in Step 6 which achieves a factor of $0.509$.
\end{theorem}

\begin{proof}

To prove the theorem, we consider a vertex $v \in V_2$ and all the edges incident to $v$ which are in $R_1 \cup R_2 \cup R_3$ (i.e. survived contention resolution on the left-hand side). To show that the required kind of CRS exists, we consider a subset of edges $S$ incident to $v$, and compute the probability that at least one of them survives as follows:
\begin{eqnarray*}
\Pr[S \cap (R_1 \cup R_2 \cup R_3) \neq \emptyset] & = & \Pr[S \cap R_1 \neq \emptyset] \\
&&+ \Pr[S \cap R_1 = \emptyset \ \& \ S \cap R_2 \neq \emptyset] \\
&&+ \Pr[S \cap (R_1 \cup R_2) = \emptyset \ \& \ S \cap R_3 \neq \emptyset].
\end{eqnarray*}

We finish the proof by establishing lower bounds on each of the three terms and applying Theorem~\ref{theorem:CRS-1}. We will find that Lemma~\ref{lem:neg-correl} is useful in the analysis of the second term.

The first case is a red edge surviving in $S$: since there can be only one red edge incident to $v$, these are disjoint events for each $(u,v) \in S$:
$$ \Pr[S \cap R_1 \neq \emptyset] = \sum_{(u,v) \in S} (1 - e^{-1/e}) x_{uv}.$$

The second case is no red edge surviving in $S$, and some blue edge surviving $S$. We divide this further into two subcases:
(a) there is only one active edge in $S$, this edge is blue, and it survives; (b) there are at least two active edges in $S$ (and hence blue), and at least one of them survives. 

The probability that $(u,v)$ is the only active edge in $S$ is $(1 - e^{-x_{uv}}) \prod_{(u',v) \in S, u' \neq u} e^{-x_{u'v}} = (e^{x_{uv}} - 1) e^{-x(S)}$ (where we use the notation $x(S)= \sum_{(u,v) \in S} x_{uv}$). The probability that this edge is red is $1 - e^{-x_{uv} / e}$, hence the probability that $(u,v)$ is the only active edge in $S$, and it is blue, is $(e^{x_{uv}} - 1) e^{-x(S)} - (1 - e^{-x_{uv} / e})$. 
Finally the probability that this edge survives is $\beta = \frac{e^{-1/e} - e^{-1}}{1 - e^{-1}}$. Hence, case (a) contributes $\sum_{(u,v) \in S} ((e^{x_{uv}} - 1) e^{-x(S)} - (1 - e^{-x_{uv} / e})) \beta$. 

The probability that there are at least two active edges in $S$ is $1 - \Pr[\mbox{no active edges } \in S] - \Pr[\mbox{exactly one active edge in } S]$, where $$\Pr[\mbox{no active edges } \in S] = e^{-x(S)}$$ and $$\Pr[\mbox{exactly one active edge in } S] = \sum_{(u,v) \in S} (e^{x_{uv}} - 1) e^{-x(S)}$$ as above. Conditioned on having at least two active (and hence blue) edges, the probability that one of them survives in $R_2$ is at least $2\beta - \beta^2$, because the survivals of two blue edges are negatively correlated due to Lemma~\ref{lem:neg-correl}. (If there are more than $2$ blue edges, we are only using two of them in this argument --- this bound could be improved if we had a correlation inequality for more than $2$ blue edges.) Hence, case (b) contributes at least $(1 -  e^{-x(S)} - \sum_{(u,v) \in S} (e^{x_{uv}} - 1) e^{-x(S)}) (2 \beta - \beta^2)$. Combining these contributions, we obtain
\begin{eqnarray*}
&&\Pr[S \cap R_1 = \emptyset \ \& \ S \cap R_2 \neq \emptyset]\\
&\geq&\sum_{(u,v) \in S} \left( (e^{x_{uv}} - 1) e^{-x(S)} - (1 - e^{-x_{uv} / e}) \right) \beta \\
&&+\left(1 -  e^{-x(S)} - \sum_{(u,v) \in S} (e^{x_{uv}} - 1) e^{-x(S)} \right) (2 \beta - \beta^2).
\end{eqnarray*}

The third case is no red or blue edges surviving in $S$ and some gray edge surviving in $S$. This case certainly happens if there are no active edges in $S$ and some gray edge survives in $S$. These two events are positively correlated, since having no active edges in $S$ can only increases the probability that gray edges are considered in Step 5, but we ignore this benefit here; we analyze the two events as independent. We have $\Pr[\mbox{no active edges in } S] = e^{-x(S)}$ and $\Pr[\mbox{no gray edge survives in }S] \leq \prod_{(u,v) \in S} (1 - \frac{1}{2e} x_{uv}^2) \leq e^{-\frac{1}{2e} \sum_{(u,v) \in S}  x_{uv}^2}$, because different gray edges incident to $v$ survive independently. Hence, this case contributes
\begin{align*}
\Pr[S \cap (R_1 \cup R_2) = \emptyset \ \& \ S \cap R_3 \neq \emptyset] &\geq e^{-x(S)} \left(1 - e^{-\frac{1}{2e} \sum_{(u,v) \in S}  x_{uv}^2} \right) \\
& \geq e^{-x(S)} (1 - e^{-\frac{1}{2e}}) \sum_{(u,v) \in S} x_{uv}^2.
\end{align*}

Combining cases the different contributions, we obtain
\begin{eqnarray*}
&&\Pr[S \cap (R_1 \cup R_2 \cup R_3) \neq \emptyset] \\ & \geq & \sum_{(u,v) \in S} (1 - e^{-1/e}) x_{uv} + \sum_{(u,v) \in S} ((e^{x_{uv}} - 1) e^{-x(S)} - (1 - e^{-x_{uv} / e})) \beta \\
  &&+  \Big( 1 -  e^{-x(S)}  - \sum_{(u,v) \in S} (e^{x_{uv}} - 1) e^{-x(S)} \Big) (2 \beta - \beta^2) \\
  &&+ e^{-x(S)} (1 - e^{-\frac{1}{2e}}) \sum_{(u,v) \in S} x_{uv}^2 \\
    & = & (1 - e^{-1/e}) x(S) - \sum_{(u,v) \in S} \Big( (1 - e^{-x_{uv}/e}) \beta + (e^{x_{uv}} - 1) e^{-x(S)} (\beta - \beta^2) \Big) \\
&&+ (1 - e^{-x(S)}) (2 \beta - \beta^2) + e^{-x(S)} (1 - e^{-\frac{1}{2e}}) \sum_{(u,v) \in S} x_{uv}^2.
\end{eqnarray*}
We simplify this expression using some elementary estimates: $e^{x_{uv}} - 1 \leq x_{uv} + (e-2) x_{uv}^2$,
and $1 - e^{-x_{uv} / e} \leq \frac{1}{e} x_{uv} + (1-1/e-e^{-1/e}) x_{uv}^2$   (for $0 \leq x_{uv} \leq 1$). 
Also, we observe that $1 - 1/e - e^{-1/e} < 0$, so we can write 
$1 - e^{-x_{uv} / e} \leq \frac{1}{e} x_{uv} + (1-1/e-e^{-1/e}) e^{-x(S)} x_{uv}^2$   (which simplifies the following bound).
We obtain
\begin{eqnarray*}
&&\Pr[S \cap (R_1 \cup R_2 \cup R_3) \neq \emptyset] \\ & \geq & (1 - e^{-1/e}) x(S) - \frac{\beta}{e} x(S) -  (\beta - \beta^2) e^{-x(S)}x(S)  +  (2 \beta - \beta^2) (1 - e^{-x(S)})\\
 && + \Big( ( e^{-1/e} - 1 + 1/e) \beta - (e-2) (\beta - \beta^2)  + (1 - e^{-\frac{1}{2e}}) \Big) e^{-x(S)} \sum_{(u,v) \in S}  x_{uv}^2.
\end{eqnarray*}
Recall that $\beta = \frac{e^{-1/e}-e^{-1}}{1-e^{-1}}$. Plugging numerical values into the last line, we find that the constant in front of $e^{-x(S)} \sum_{(u,v) \in S} x_{uv}^2$ is positive ($\simeq 0.019$). Hence we can ignore the last line and write
\begin{eqnarray*}
&&\Pr[S \cap (R_1 \cup R_2 \cup R_3) \neq \emptyset] \\ & \geq & (1 - e^{-1/e} - \beta/e) x(S) - e^{-x(S)} x(S) (\beta - \beta^2) + (1 - e^{-x(S)}) (2 \beta - \beta^2).
\end{eqnarray*}
It can be verified that this is a concave function of $x(S) \in [0,1]$, which is lower-bounded by $\gamma \, x(S)$ where $\gamma$ is the value of the right-hand side when $x(S)=1$,
$$ \gamma =  (1 - e^{-1/e}) - e^{-1} (4\beta - 2 \beta^2) + (2\beta - \beta^2) \geq 0.509.$$
Hence, by Theorem~\ref{theorem:CRS-1}, there is a CRS which preserves each edge with probability at least $\gamma x_{uv} \geq 0.509 x_{uv}$.
\end{proof}

\section{An application to a combinatorial allocation problem}
\label{sec:application}

In this section, we consider a problem that can arise in a scheduling setting. Consider a set of customers $X$ each of whom needs to be scheduled to attend one of $m$ possible events, over a time period of $n$ hours, each event one hour long. Naturally, each customer can attend only one event at a time, and wants to attend each event at most once. What is the ``best" possible way to schedule the customers?

One way to formalize the problem is as follows:

\paragraph{Allocation with row/column disjointness constraints.} 
\

Given a finite set $X$ and valuations $v_{st}:2^X \rightarrow \RR_+$ for $s \in [m], t \in [n]$, choose sets $S_{st} \subseteq X, s \in [m], t \in [n]$ such that for every $s$, $\{S_{st}: t \in [n] \}$ are disjoint and for every $t$, $\{S_{st}: s \in [m] \}$ are disjoint, in order to maximize $\sum_{s,t} v_{st}(S_{st})$.

\

Here, we are thinking of the $s$ index as denoting the events, the $t$ index as denoting time, and then $S_{st}$ is the set of customers attending event $s$ at time $t$. The welfare of a set of customers $S$ attending event $s$ at time $t$ is given by $v_{st}(S)$. The goal is maximize the total welfare of a schedule.

We note that this is a special case of maximization subject to a bipartite matching constraint, due to the following reduction: We define a bipartite ``conflict graph'' $G_a \subset K_{m,n}$ for each $a \in X$. The interpretation of using edge $(s,t)$ in $G_a$ is that we include element $a$ in $S_{st}$. Since the sets $S_{st}$ should be disjoint for each fixed $s$ and each fixed $t$, this corresponds to the constraint that we should choose a matching in each graph $G_a$. The objective function is $\sum_{s,t} v_{st}(S_{st})$ where $S_{st} = \{a: \mbox{we choose edge }(s,t) \mbox{ in } G_a\}$. 
Hence, this is a maximization problem over matchings in the disjoint union of the graphs $G_a, a \in X$.

If the valuation functions $v_{ij}$ are monotone submodular, this is a monotone submodular maximization problem over bipartite matchings, for which there is a known $(1/2-\epsilon)$-approximation due to \cite{LSV10}.
Here we show an improved approximation for the above allocation problem, in fact even for the larger class of fractionally subadditive valuations~\footnote{A function $v:2^X \to \RR_+$ is fractionally subadditive, if 
for any $S \subseteq X$ and a fractional cover $\sum_i \alpha_i \b1_{T_i} \geq \b1_S$, $\alpha_i \geq 0$, we have
$\sum_i \alpha_i v(T_i) \geq v(S)$. It is known that every submodular function is fractionally subadditive.}
(but not for submodular maximization over matchings), in the demand oracle model.

\begin{theorem}
For fractionally subadditive valuations $v_{st}$ given by demand oracles (returning $\argmax_{S \subseteq X}$ $v_{st}(S) - \sum_{a \in S} p_a$ for given prices $p_a$), there is a $0.509$-approximation for the problem of Allocation with row/column disjointness constraints.
\end{theorem}

\begin{proof}
We use the ``Configuration LP'' which is a standard tool for allocation problems of this type:
\begin{eqnarray*}
\max \sum_{s=1}^{m} \sum_{t=1}^{n} \sum_{S \subseteq X} v_{st}(S) x_{s,t,S} \\
\forall a \in X; \forall s \in [m]; \sum_{t=1}^{n} \sum_{S: a \in S} x_{s,t,S} \leq 1 \\
\forall a \in X; \forall t \in [n]; \sum_{s=1}^{m} \sum_{S: a \in S} x_{s,t,S} \leq 1 \\
\forall s,t \in [n]; \sum_S x_{s,t,S} = 1 \\
\forall s,t \in [n]; \forall S; x_{s,t,S} \geq 0. \\
\end{eqnarray*}
This LP is exponentially large but it can be solved in polynomial time using demand oracles for $v_{st}$, since a separation oracle for the dual LP is exactly a sequence of calls to the demand oracles for $v_{st}$. This is very similar e.g.~to the approach in \cite{FV10}.

Given a fractional solution $x_{s,t,S}$, we can round it as follows:
For each $s,t$ independently, sample a random set $R_{st}$ which is equal to $S$ with probability $x_{s,t,S}$. For each item $a \in X$, form a bipartite conflict graph $G_a$ where edge $(s,t)$ appears if $a \in R_{st}$. This is a random graph where edge $(s,t)$ appears with probability $\sum_{S: a \in S} x_{s,t,S}$ and hence due to our constraints this is in expectation a fractional matching in $G_a$.

Using our CRS for bipartite matchings, we can select a matching $M_a \subseteq G_a$, such that $\Pr[(s,t) \in M_a \mid (s,t) \in G_a] \geq 0.509$. Finally, we define sets $S_{st} = \{ a \in X: (s,t) \in M_a \}$, which form our solution. Observe that $S_{st} \subseteq R_{st}$ and each element of $R_{st}$ survives in $S_{st}$ with conditional probability at least $0.509$ (the survivals of different elements are correlated, but this is not a problem). By the defining property of fractionally subadditive functions (see e.g. Lemma 1.7 in \cite{FV10}),
$$ \E[v_{st}(S_{st})] \geq 0.509 \cdot \E[v_{st}(R_{st})] = 0.509 \sum_S v_{st}(S) x_{s,t,S}.$$
\end{proof}

In our analysis, we do not require the monotonicity of the contention resolution scheme, just like it is not required in \cite{FV10}.

\backmatter

\bmhead{Acknowledgments}

We would like to thank Chandra Chekuri for stimulating discussions. The second author is supported by NSF Award 2127781.

\section*{Declarations}

\begin{itemize}
\item Funding: The authors are supported by NSF Award 2127781.
\item Conflict of interest/Competing interests: None.
\item Ethics approval: Not applicable.
\item Consent to participate: Not applicable.
\item Consent for publication: Not applicable.
\item Availability of data and materials: Not applicable.
\item Code availability: Not applicable.
\item Authors' contributions: The authors contributed equally to this paper.
\end{itemize}

\bibliography{ref}% common bib file

\end{document}